\documentclass[twocolumn,showpacs,showkeys,amsmath,amssymb,aps,prl]
{revtex4-1}
\usepackage{float}
\usepackage[final]{graphicx} 
\usepackage{dcolumn}
\usepackage{bm}
\usepackage[dvips]{color}

\begin{document}

\title{Streamwise-localized solutions at the onset of turbulence in pipe flow}

\author{M. Avila$^1$, F. Mellibovsky$^2$, N. Roland$^3$ \& B. Hof$^{3,4}$}

\affiliation{$^1$Institute of Fluid Mechanics,
  Friedrich-Alexander-Universit\"at Erlangen-N\"urnberg, 91058
  Erlangen, Germany\\ $^2$Castelldefels School of Telecom and
  Aerospace Engineering (EETAC), Universitat Polit\`ecnica de
  Catalunya, 08860 Barcelona, Spain\\ $^3$Max Planck Institute for
  Dynamics and Self-Organization (MPIDS), 37077 G\"ottingen,
  Germany\\ $^4$Institute of Science and Technology Austria, 3400
  Klosterneuburg, Austria }

\date{\today}

\begin{abstract}
  Although the equations governing fluid flow are well known, there
  are no analytical expressions that describe the complexity of
  turbulent motion.  A recent proposition is that in analogy to low
  dimensional chaotic systems, turbulence is organized around unstable
  solutions of the governing equations which provide the building
  blocks of the disordered dynamics. We report the discovery of
  periodic solutions which just like intermittent turbulence are
  spatially localized and show that turbulent transients arise from
  one such solution branch.
\end{abstract}


\maketitle 

Fluids move in a well ordered fashion (laminar flow) when their
velocity is small and in this case the flow field can usually be
analytically derived from the equations of motion, the Navier--Stokes
equations. However, as the inherent velocity and length scales become
large, turbulence sets in and most flows of practical interest are
highly disordered in space and time. Landau and Hopf proposed in the
forties that this transition occurs via an infinite sequence of
bifurcations starting from laminar flow
\cite{landau1944,*hopf1948}. This route to turbulence, later shown to
consist of only a few bifurcations by Ruelle and Takens
\cite{ruelle1971,*landau1987}, is a well established paradigm for
transition in many systems. In flow down a straight circular pipe,
however, turbulence arises despite linear stability of the laminar
flow \cite{reynolds1883}, and thus the former scenario is in principle
inapplicable. Moreover, in pipes just like in many other wall-bounded
flows turbulence first manifests itself in localized spots surrounded
by laminar flow.  Although experimental observations of localized
turbulent structures date back to the first comprehensive
investigations of turbulence \cite{reynolds1883} and their structure
and kinematics have been studied extensively
\cite{wygnanski1973,*wygnanski1975,*bandyopadhyay1986,*nishi2008,*shimizu2009,*vanDoorne2009,*hof2010},
a theoretical understanding is missing. More recent studies have shown
that turbulent spots (called puffs in pipe flow) are generally of
transient nature and that their decay is memoryless
\cite{hof2006,*hof2008,avila2010}. Nevertheless turbulence eventually
becomes sustained once these structures begin to proliferate and their
spreading rate outweighs their decay \cite{avila2011}. The Reynolds
number ($Re=DU/\nu$, where $D$ is the pipe diameter, $U$ the mean
velocity and $\nu$ the kinematic viscosity of the fluid) at which
these processes balance marks a phase transition to sustained
turbulence. Despite such recent advances, how these turbulent
structures arise from the equations of motion is unknown.

Numerical studies of flows in short periodic domains led to the
important discovery of invariant solutions of the Navier--Stokes
equations featuring the main ingredients of the self-sustaining cycle
of turbulent shear flows \cite{nagata1990,*waleffe1998}. In pipe flow,
the simplest of these solutions are traveling waves
\cite{faisst2003,*wedin2004}, satisfying
\begin{equation}
 \boldsymbol v(x,r,\theta,t) = \boldsymbol v(x-ct,r,\theta),
\end{equation}
where $(x,r,\theta)$ are cylindrical coordinates, $t$ time and $c$ the
wave-speed. Traveling waves are frozen as they propagate, i.e. they
are relative equilibria. Although all these exact numerical solutions
are unstable, and hence cannot be directly observed in experiments,
the number of unstable directions is small, so it is expected that
they play an important role in organizing the phase-space dynamics of
turbulence
\cite{kerswell2005,*eckhardt2007,*gibson2008,*kawahara2012}. As
traveling waves have no dynamics but only drift in the propagation
direction, more complex solutions are required to capture the
properties of turbulent flows. The next level of complexity in the
hierarchy of invariant solutions of the governing equations is
provided by relative periodic orbits (RPOs)
\begin{equation}
\boldsymbol v(x,r,\theta,t) = \boldsymbol
v(x-\overline{c}T,r,\theta,t+T),
\end{equation} 
for which the motion appears as $T$-periodic in a frame co-moving at
speed $\overline{c}$. Relative periodic orbits bifurcating from
traveling waves \cite{duguet2008b} and embedded in turbulence
\cite{willis2013} have been recently discovered in short pipes.

Although some aspects of the traveling wave solutions found in small
domains, like the symmetry and the vortex streak arrangement have also
been observed in turbulent pipe experiments
\cite{hof2004,*hof2005,*delozar2012}, the streamwise structure is
qualitatively different. While traveling waves are streamwise
periodic, with a periodicity of a few $D$, all turbulent structures
observed close to onset are localized. Turbulent puffs have distinct
laminar-turbulent interfaces characterized by a sharp velocity change
at the upstream interface and a slow adjustment downstream. In this
\emph{Letter}, we present the first localized solutions that contain
all spatial features of turbulent puffs and show how turbulent
transients emerge from them.

Numerical simulations of pipe flow were carried out using a spectral
code \cite{meseguer2007} and a hybrid spectral finite-difference code
\cite{willis2009}, with excellent agreement between them. The
computational domain was chosen to be long ($40D$) with periodic
boundary conditions in the streamwise direction. In such long domains,
just like in experiments, turbulence takes the form of localized puffs
and the agreement with experiments even of very subtle features like
lifetime statistics is very good \cite{avila2010}.

At first our investigation focused on the laminar-turbulent
phase-space boundary by looking for initial conditions that neither
turn turbulent nor relaminarise but remain in the dividing edge
\cite{schneider2007}. In long pipes the attractor in the edge (called
edge state) was found to be localized but at the same time chaotic
\cite{mellibovsky2009,willis2009} and the dynamics turned out to be
too complex to identify underlying invariant solutions. Although
approaches to nearly periodic dynamics were reported in studies of
symmetric invariant subspaces, in long pipes the edge state was always
found to be chaotic \cite{duguet2010}. Here we simplified the problem
by restricting the dynamics subject to a $\pi$-rotational symmetry
with respect to the pipe axis
\begin{equation}
  [u,v,w](x,r,\theta,t) = [u,v,w](x,r,\theta+\pi,t)
\end{equation}
and the reflectional symmetry 
\begin{equation}\label{eq:RS}
  [u,v,w](x,r,\theta,t) = [u,v,-w](x,r,-\theta,t),
\end{equation}
where $u$, $v$ and $w$ are the axial, radial and azimuthal velocities,
respectively. The reflectional symmetry \eqref{eq:RS} prohibits
rotations about the pipe axis. Note that any solutions found in the
subspace are necessarily also solutions of the full space and hence
represent physical (symmetric) flow states.

\begin{figure}
  \centering
  \includegraphics[width=0.95\linewidth,clip=]{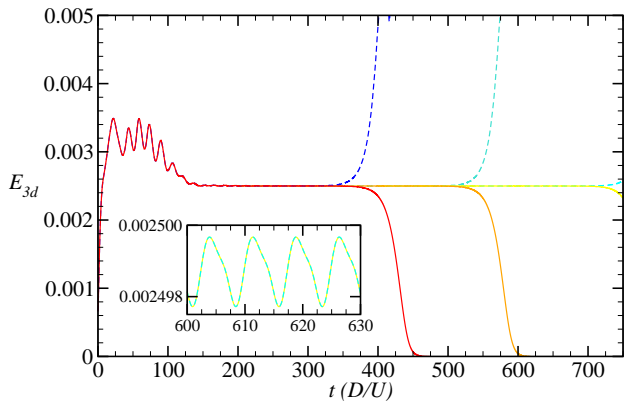}
  \caption{Dynamics of pipe flow at the edge ($Re=2200$). At $t=0$ a
    disturbance is applied to the laminar flow and the evolution of
    kinetic energy (of three-dimensional Fourier modes) is
    subsequently monitored. The dashed lines correspond to flow
    trajectories that shoot up to turbulence, whereas the solid lines
    show trajectories that relaminarize. The edge-tracking algorithm
    is applied to obtain trajectories that hang around on the edge of
    chaos. 
    The periodic oscillations shown in the inset (close up) suggest that
    trajectories on the edge are attracted to a RPO.}
  \label{fig:edge}
\end{figure}

The edge-tracking algorithm is as follows. First a localized
disturbance is applied to the laminar flow \cite{mellibovsky2007} and
if sufficiently strong it evolves into a turbulent puff. Subsequently,
the amplitude of this puff, to which the laminar parabolic flow has
been subtracted, is rescaled to obtain a new initial condition
$\boldsymbol v_{\alpha}=\boldsymbol v_\text{lam} + \alpha(\boldsymbol
v-\boldsymbol v_\text{lam})$, where $\alpha$ is a constant
$\alpha\in(0,1)$, $\boldsymbol v$ the velocity field of the puff and
$\boldsymbol v_\text{lam}$ the laminar flow.  A simple bisection
algorithm is then used to find the value of $\alpha$ for which the
temporal evolution of $\boldsymbol v_{\alpha}$ neither relaminarizes
nor goes to turbulence but remains on the edge. The procedure is
illustrated in figure~\ref{fig:edge} at $Re=2200$. After an initial
transient the temporal evolution rapidly relaxes onto a periodic
oscillation, suggesting that the edge state is a RPO. Note that as
time evolves new refinement bisection iterations have to be applied to
keep the trajectory on the edge.

\begin{figure}
  \centering
  \begin{tabular}{l}
    \includegraphics[width=0.95\linewidth]{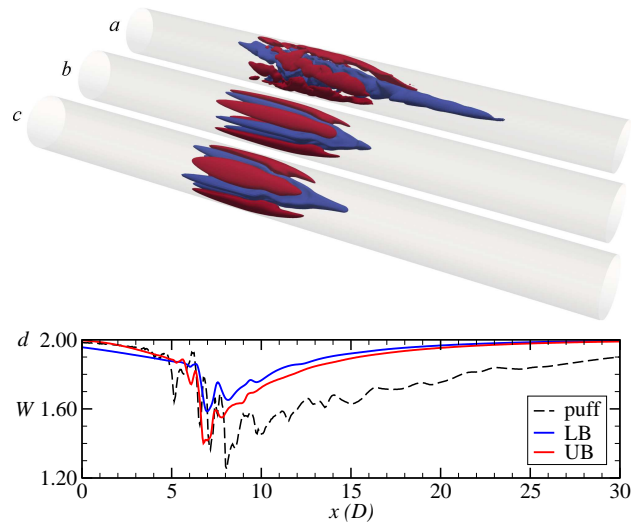}\\
    \includegraphics[width=0.95\linewidth,clip=]{fig2d.eps} 
  \end{tabular}
  \caption{(a) Turbulent puff at $Re=1900$ and reflection-symmetric
    RPOs with $\pi$-rotational-symmetry at: (b) the edge (LB) at
    $Re=1900$ and (c) UB at $Re=1500$. Isosurfaces of streamwise
    velocity at $0.2U$ (red) and $-0.2U$ (blue) are shown. The laminar
    profile has been subtracted in all cases to highlight the
    three-dimensional structure of the flow and the views have been
    shrank by a factor of 4 in the streamwise direction. $40D$ are
    shown out of a simulation domain of $50D$ (puff) and $40D$ (LB,
    UB). (d) Streamwise velocity along the pipe centerline for the
    structures in (a)--(c).}
  \label{fig:iso3d}
\end{figure}

A snapshot of the edge velocity field was fed as initial guess into a
purposely designed Newton--Krylov solver based on the time-stepping
code \cite{meseguer2007} using standard techniques
\cite{wiswanath2006,duguet2008b} and rapidly converged to a RPO with
period $T=15.0\,D/U$ and average drift speed
$\overline{c}=1.52\,U$. Note that in order to achieve convergence we
require that the residual $r=||\boldsymbol v(T)-\boldsymbol
v(0)||<10^{-10}||\boldsymbol v(0)||$, where the velocity field
$\boldsymbol v(T)$ has been appropriately shifted to account for
drift.  Figure~\ref{fig:edge} shows that the energy oscillations have
a period of $T/2$. This is due to a spatiotemporal symmetry possessed
by this solution: at $t=T/2$ the velocity field is the same as $t=0$
but reflected with respect to the plane at $\theta=45^\circ$ (note
that the plane of imposed reflection-symmetry is at
$\theta=0$). Figure~\ref{fig:iso3d}b shows a snapshot of the RPO. The
similarity in the topology of its low and high velocity streaks with
those of a turbulent puff (shown in~\ref{fig:iso3d}a) is remarkable.
A close inspection of the topology of streaks and vortices of this
solution points at a possible connection with a stream-wise periodic
traveling wave \cite{duguet2008} (D2).  We found that at $Re=2200$
this traveling wave is the edge state in short pipes of length
$\lambda\lesssim 5D$, whereas in the range $5\lesssim\lambda\lesssim
10D$ the edge state is chaotic. Although a localized RPO is
  obtained as long as $\lambda\gtrsim 10D$, for $\lambda\lesssim 15D$
  the periodic boundary conditions ostensibly interfere with
  localization. For the pipe length $\lambda=40D$ used in the results
  presented here, the periodic boundary conditions have no longer an
  effect on localization (which was tested by repeating some
  simulations for $\lambda=80D$).

As the Reynolds number is reduced the localized RPO (henceforth
referred to as LB, which stands for lower branch solution) keeps
fulfilling its role of separating trajectories that relaminarize from
those that increase in energy towards turbulence. For $Re<1530$
trajectories above the edge no longer result in turbulent transients
but approach instead a stable (within the $\pi$-rotational- and
reflection-symmetric space) localized RPO. The visualisation of this
new solution (hereafter UB, standing for upper branch solution) is
shown in figure~\ref{fig:iso3d}c and reveals a striking structural
resemblance to turbulent puffs. As pointed out above, a typical
signature of puffs is the sharp transition from laminar to turbulent
flow at the trailing interface followed by a slow recovery towards the
laminar velocity along its diffuse leading interface (see the black
curve in figure~\ref{fig:iso3d}d). This landmark of puffs is shared by
LB (red curve) and UB (blue curve) and further demonstrates that the
properties of localized turbulence can be captured by exact numerical
solutions of the Navier--Stokes equations.

\begin{figure}
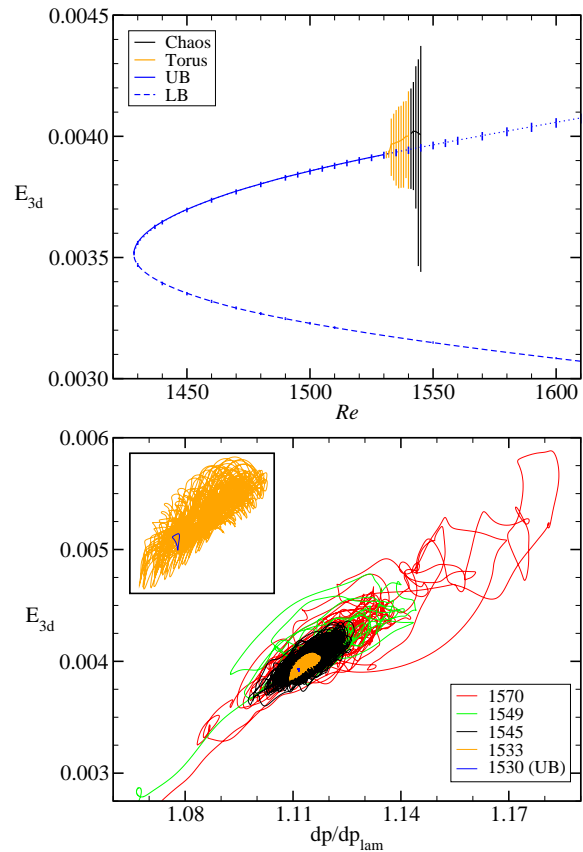

  \centering
  \begin{tabular}{r}
    \includegraphics[scale=0.32,clip=]{fig3a.eps}\\
    \includegraphics[scale=0.32,clip=]{fig3b.eps}
  \end{tabular}
  \caption{(a) A saddle-node bifurcation gives rise to localized RPOs
    at $Re\approx 1430$: UB is stable up to $Re\approx 1530$, where it
    undergoes a supercritical Neimark-Sacker bifurcation leading to a
    relative 2-torus. Subsequently the torus breaks up to chaos at
    $Re\approx 1540$ and the chaos becomes transient at $Re\approx
    1545$. The bars show the variation of energy over a period
    (Newton-converged LB and UB) and over long runs (torus and
    chaos). LB has a single unstable direction and is the edge
    state. (b) Phase-portrait of the dynamics at several $Re$
    projected onto a two-dimensional plane defined by the energy (of
    three-dimensional Fourier modes) and pressure gradient required to
    drive a constant flow rate, normalized with the pressure gradient
    of laminar flow. The inset is a close up showing the UB and
    torus.}
  \label{fig:bifdiag}
\end{figure}

At $Re \approx 1430$ UB merges with LB at a saddle-node bifurcation
(see figure~\ref{fig:bifdiag}a) and below this bifurcation no dynamics
other than laminar are found. By continuing the UB towards larger
Reynolds number we could identify a bifurcation cascade leading to
turbulent transients. At $Re\approx 1530$ the UB undergoes a
Neimark--Sacker bifurcation leading to a stable 2-torus that breaks up
into chaos at $Re\approx 1540$. Although at the onset of chaos the
attractor explores only a small portion of the phase space, this
portion grows explosively as $Re$ is increased and the chaotic
attractor appears to collide with LB at $Re\approx 1545$. This
boundary crisis is likely related to the appearance of a homoclinic
tangle on the edge \cite{vanVeen2011,*lebovitz2012}. Beyond this point
the attractor becomes leaky: trajectories can relaminarize after long
transients. Following the ensuing chaotic saddle to larger $Re$
confirms that turbulence in the subspace originates at this
bifurcation, as illustrated in the phase-space portrait of
figure~\ref{fig:bifdiag}b. We note that similar bifurcation scenarios
but starting from relative equilibria have been observed in short
pipes \cite{mellibovsky2012} and in small plane Couette cells
\cite{kreilos2012}, thus lacking the spatial complexity and laminar
turbulent interfaces observed in practice. In these small cells a
chaotic attractor emerges via period doubling bifurcations and
subsequently leads to transients \cite{kreilos2012}.

The robustness of the RPOs and transition scenario were tested with
respect to spatial resolution and time-step $\delta t$. We used
$\delta t=0.0025D/U$ and $K=\pm 320$ axial Fourier modes, $M=\pm 16$
azimuthal Fourier modes (for $\theta\in[0,\pi]$) and $N=40$ points in
the radial direction. With these values the solutions are well
converged and the bifurcation points are accurate to better than
$0.5\%$. For lower resolutions the bifurcations are shifted towards
lower $Re$, whereas the opposite effect is observed by increasing
$\delta t$. Nevertheless, the scenario remains qualitatively
unchanged. Note that in the full space the solutions found here have
several additional instabilities and hence cannot be computed by edge
tracking and time-stepping. We performed several simulations starting
from them but dropping the symmetry restrictions and observed similar
transients.

In summary, we have discovered exact numerical solutions of the
Navier--Stokes equations that share structure and spatial complexity
with turbulence at onset. We have furthermore shown that a bifurcation
sequence is responsible for giving rise to transient turbulence. In
contrast to the classical Ruelle--Takens model, in pipe flow chaotic
motion arises locally originating from the discovered localized
solutions. This is a key difference to the much simpler transition
scenarios in linearly unstable flows, such as Rayleigh--B\'enard
convection \cite{libchaber1982} and Taylor--Couette flow
\cite{gollub1975}, where the bifurcation sequence starts from the base
flow and instability occurs globally in space. Localized solutions can
therefore be regarded as the nuclei of disordered motion in linearly
stable shear flows.  It is likely that in full space chaotic dynamics
simultaneously arises from distinct nuclei and that the corresponding
repellers merge in global bifurcations as $Re$ grows, increasing the
complexity of the turbulent transients. One of the outstanding
challenges towards an understanding of the spatiotemporal complexity
encountered in shear flows close to onset is the identification of the
mechanism leading to spatial localization \cite{knobloch2008,*schneider2010b}.

\begin{acknowledgments}
We thank A.~P.~Willis for sharing his code. The research leading to
these results has received funding from the Max Planck Society and the
European Research Council under the European Union's Seventh Framework
Programme (FP/2007-2013) / ERC Grant Agreement 306589. We acknowledge
computing resources from GWDG and the J\"ulich Supercomputing Centre
(grant HGU16).
\end{acknowledgments}

\bibliography{localized}


\end{document}